\documentstyle[version2,preprint,aps]{revtex} 
\begin{document}
\draft 		
\begin{title}
Symmetry dependence of phonon lineshapes \\
in superconductors with anisotropic gaps
\end{title}
\author{T. P. Devereaux}
\begin{instit}
Department of Physics\\
University of California\\
Davis, CA 95616\\
\end{instit}
\begin{abstract}
	The temperature dependence below $T_{c}$ of the lineshape of optical
phonons of different symmetry as seen in Raman scattering is
investigated for superconductors with anisotropic energy gaps. It is shown
that the symmetry of the electron-phonon vertex produces non-trivial
couplings to an
anisotropic energy gap which leads to unique changes in the phonon lineshape
for phonons of different symmetry.  The phonon lineshape is calculated
in detail for $B_{1g}$ and $A_{1g}$ phonons in a superconductor with
$d_{x^{2}-y^{2}}$ pairing symmetry. The role of satellites peaks
generated by the electron-phonon coupling are also addressed.
The theory accounts for the substantial
phonon narrowing of the $B_{1g}$ phonon, while
narrowing of the $A_{1g}$ phonon
which is indistinguishable from the normal state is shown,
in agreement with recent measurements on Bi$_{2}$Sr$_{2}$CaCu$_{2}$O$_{8}$
and YBa$_{2}$Cu$_{3}$O$_{7}$.
\end{abstract}
\vspace{0.5 cm}
\narrowtext
\newpage
\section{Introduction}
Optical phonons observed via Raman scattering have provided a large amount
of information concerning the energy gap in high-T$_{c}$ superconductors
\cite{cardona}, and there have been attempts to describe
the changes in the phonon lineshapes below T$_{c}$ in $s-$wave \cite{ZZ}
and $d-$wave superconductors \cite{nicol}.  It is believed that the changes
in the phonon lineshape below T$_{c}$ are due in part to changes in the phonon
self-energy resulting from coupling between phonons and quasiparticles. It has
been argued that if the optical phonon has a frequency below the pair
threshold energy $2\Delta$, then the phonon's
linewidth decreases (narrows)
and its frequency renormalizes to lower frequencies
(softens) as the quasiparticles
become frozen out. However, for a phonon near $2\Delta$, the linewidth is
predicted to grow due to the enhancement of the density of states at the
gap edge and there can be either pronounced phonon softening or hardening
depending on which side of the threshold the phonon is located.
This simple
picture has been employed to determine the position of $2\Delta$ in
the cuprate superconductors.

	However, this simple analysis applied to the cuprate systems
has revealed that the above picture
is a bit misleading.   The above scenario has yielded a value for the
energy gap that is different for different types of optical phonons and is
thus symmetry-dependent. For the case of the Bi 2:2:1:2
system, where very clean surfaces can be obtained, a low frequency
phonon which transforms according
to $A_{1g}$ symmetry (located at 464 cm$^{-1}$, connected with the
bridging Oxygen vibrations) shows a downward
frequency shift (softening) while no substantial linewidth change
from the normal state can be resolved from the data\cite{burns}.
However, the $B_{1g}$ phonon (285 cm$^{-1}$, connected with the
antisymmetric out of plane O(2) and O(3) vibrations in the Cu-O
plane) on the contrary
shows a small frequency softening but a substantial linewidth
narrowing below T$_{c}$\cite{leach}.
Similar behavior is seen in the YBCO systems\cite{irwin}, where such
a large difference in behavior between the $A_{1g}$ and $B_{1g}$ phonons
in part led the authors of Ref. \cite{irwin}
to suggest that these two phonons interact with different electronic systems.

There has been no satisfactory theoretical explanation for the behavior of
the different phonons.
The main problem in addressing these experiments with the existing
theories concerns the lack of attention
paid to the symmetry dependence of the optical phonons. However, this
symmetry dependence can be an important tool to uniquely determine the
${\bf \hat k}-$dependence of the energy gap around the Fermi surface.
It has been shown that the electronic
contribution to Raman scattering can provide a large amount of polarization
(symmetry) dependent information that allows for a stringent test to made
to determine the actual symmetry of the energy gap in superconductors\cite{us}.
It was shown that the coupling between the Raman vertex and an anisotropic
gap leads to symmetry dependent spectra, with peak positions and low frequency
and temperature behavior dependent on polarization orientations.  These
changes in the spectra allow for a direct determination of
$\mid\Delta({\bf \hat k})\mid$.
Good agreement with the electronic Raman spectra
taken on very clean BSCCO surfaces was obtained using a gap which was
predominantly or entirely of $d_{x^{2}-y^{2}}$ symmetry, where the peak
position and the low frequency behavior of the spectra could be
straightforwardly accounted for.  The symmetry dependence of the data led
to the conclusion that the gap must be predominantly of $B_{1g}$ character.
Since the phonon self-energy is very similar to the electronic Raman density
response, the same type of analysis for the electronic contribution to Raman
scattering can be made to the phonons as well, leading to a further check
on the predictions recently made concerning the energy gap in the cuprate
materials.

	We propose an alternative explanation for the symmetry
dependence of the Raman shifts based upon nontrivial couplings of
phonons of different symmetry with an anisotropic energy gap.
Close attention will be paid to the role of the
electron-phonon vertex, and consequences of its {\bf k}-dependence
will be addressed.
Most importantly, it is shown that the lineshape is polarization dependent
for anisotropic superconductors
and different dependences on temperature can be used to determine not
only the magnitude but the symmetry dependence of the energy gap.
Moreover, it is shown that the peak of the self energy can be
located at frequencies {\it below} $2\Delta_{max}$ for certain polarizations
which have an symmetry orthogonal to that of the energy gap. Thus if the
symmetry of the phonon is neglected, values of the energy gap inferred from
changes in the phonon lineshape using an isotropic s-wave theory
will be underestimated.
In particular, the phonon spectral function for a
superconductor with $d_{x^{2}-y^{2}}$ symmetry is examined and a
comparison is made with experimental data on both the $B_{1g}$ and the
$A_{1g}$ phonons in BSCCO and YBCO.
It is shown that satisfactory agreement can be
obtained which reconciles the differences between the
$A_{1g}$ and $B_{1g}$ phonon lineshapes.

\section{Phonon spectral function}

The phonon spectral function is given by
\begin{equation}
Im
D(\omega)={4\omega_{0}^{2}\Sigma^{\prime\prime}(\omega)\over{[\omega^{2}-\omega_{0}^{2}-2\omega_{0}\Sigma^{\prime}(\omega)]^{2}+4\omega_{0}^{2}\Sigma^{\prime\prime 2}}},
\end{equation}
where $\omega_{0}$ is the optical phonon frequency and
$\Sigma^{\prime},\Sigma^{\prime\prime}$ are the real and imaginary parts of the
phonon self-energy, respectively. The real part of the self
energy renormalizes the position of the phonon, while the imaginary part
governs the linewidth.  The interaction of optical
phonons and electrons can be simply written as
\begin{equation}
H_{e-ph}=\sum_{{\bf k,q},\gamma,\sigma}
g_{{\bf k}}^{\gamma}({\bf q})c^{\dagger}_{{\bf k-q},\sigma} c_{{\bf k},\sigma}
(b^{\dagger}_{{\bf q},\gamma} +b_{{\bf q},\gamma}),
\end{equation}
where $g_{\bf k}^{\gamma}({\bf q})$
is the matrix element for scattering an electron
from ${\bf k} \rightarrow {\bf k-q}$, and
$b_{{\bf q},\gamma}, b^{\dagger}_{{\bf q},\gamma}$
are the field operators for phonons of branch
$\gamma$. The details of the scattering matrix elements depend on the nature
of the mechanism of the electron-phonon coupling and the symmetry of
the lattice vibration.  In this paper we only consider the symmetry of
the matrix element and leave a treament of the mechanism and magnitude of
the coupling for future consideration\cite{hung}.

We take the ${\bf k}$ dependence along the
Fermi surface of the vertex into account by
expanding in terms of Fermi surface harmonics $\Phi$ for small ${\bf q}$,
\begin{equation}
g_{\bf k}^{\gamma}=\sum_{L}g_{L}^{\gamma}\Phi_{L}^{\gamma}({\bf \hat k}),
\end{equation}
where the index $L$ indicates the order of polynomial that transforms
according to the $\gamma -th$ representation of the point group of the
crystal. For cylindrical Fermi surfaces, $L$ can be replaced by azimuthal
quantum numbers.
The symmetry of the optical phonon enters into
the matrix elements $g_{{\bf k}}^{\gamma}$.
The matrix elements for the phonons accessible to in-plane polarizations
are given for a cylindrical (2-D) Fermi surface in terms of azimuthal
angle $\phi$ as
\begin{eqnarray}
g^{A_{1g}}_{\bf \hat k}&=&g_{L=0}^{A_{1g}}+g_{L=4}^{A_{1g}}\sqrt{2}
\cos(4\phi)+\dots
\nonumber \\
g^{B_{1g}}_{\bf \hat k}&=&g_{L=2}^{B_{1g}}\sqrt{2} \cos(2\phi)+\dots
\end{eqnarray}
where we have dropped higher order terms, arguing that they are more
anisotropic than the terms considering here and will hence be of minor
importance.
The $L=2$ term for the $A_{1g}$ channel which is present for z dispersion
is absent here and the $L=4$ term is the first anisotropic
term in the series in this case\cite{us}.
Also, since there is no dispersion in the z direction in this case, there are
no contributions to the $E_{g}$ channels.  Consequences of the Fermi surface
and the resulting response functions are considered in a forthcoming
publication\cite{further}, and thus for our purposes we will confine
our attention to only cylindrical Fermi surfaces.

The form of the $e-ph$ interaction Eq. (2) is similar to the electronic
contribution to Raman scattering in the case of non-resonant scattering with
the replacement of the effective Raman vertex by the $e-ph$ coupling vertex.
Thus we can proceed along the lines recently taken for the case of the
electronic Raman scattering\cite{us}, where it was shown that the
Raman response is extremely polarization dependent for superconductors with
an anisotropic energy gap. Moreover, it was shown
that the collective modes which appear in the case of $d-$wave superconductors
are of little importance to the Raman response\cite{us,further}.
We can then separate the self energy in two parts
\begin{equation}
\Sigma({\bf q}, \omega)=\Sigma({\bf q}=0,\omega) +
\delta \Sigma({\bf q},\omega),
\end{equation}
Delaying a discussion of $\delta \Sigma$ until Section III,
we can write down the spectrum of the self energy at $q=0$ in the
pair approximation, e.g., neglecting collective modes as
\begin{equation}
\Sigma^{\prime\prime}_{g,g}({\bf q=0},\omega)= -{4N_{F}\over{\omega}}\langle
{\mid g_{{\bf \hat k}}^{\gamma}\mid^{2}
\mid\Delta({\bf \hat k})\mid^{2}\Theta(\omega^{2}-4\mid\Delta({\bf \hat
k})\mid^{2})\over{\sqrt{\omega^{2}-
4\mid\Delta({\bf \hat k})\mid^{2}}}}\rangle \tanh(\omega/4T).
\end{equation}
The subscript $g,g$ denotes the pair susceptibility calculated with
vertices $g$.
The real part can be obtained via a Kramers-Kronig transformation.
Here $\langle \dots \rangle$ denotes an average over the Fermi surface,
$N_{F}$ is the density of states per spin at the Fermi level, $\Theta$
is a Theta function and $\Delta({\bf \hat k})$ is the generalized
k-dependent energy gap.
We see that if the gap is
isotropic, ($\Delta({\bf \hat k})=\Delta$),
the average around the Fermi surface is frequency independent and thus
the symmetry of the vertex only determines an overall prefactor of the
self energy.  Also, since the imaginary part of the self-energy has
a divergence at the pair threshold energy $2\Delta$ a phonon with a frequency
below the threshold should be infinitely sharp (neglecting strong-coupling
effects).
However, if the gap is anisotropic, the vertex and gap couple
when averaging over the Fermi surface to produce non-trivial changes in
the self-energy of phonons of different symmetries. Further, if the gap
vanishes on the Fermi surface, the presence of the nodes can provide
decay channels for the phonon leading to a finite linewidth for all
non-zero frequencies\cite{nicol}.

The isotropic ($L=0$)
density-like terms will be coupled to the long range Coulomb forces and
thus we must take screening of the vertex into account.  Summing R.P.A.
diagrams we recover the known result at $q=0$,
\begin{equation}
\Sigma^{sc.}=\Sigma_{g,g}-\Sigma_{g,1}^{2}/\Sigma_{1,1},
\end{equation}
where $1$ denotes the $L=0$ contribution of the vertex $g$\cite{me,kandd}.
Therefore we see that the $L=0$ terms are
completely screened for ${\bf q}=0$ as a consequence of the long range
Coulomb interactions and do not contribute to the Raman response.

	Carrying out the integrations in Eq. (6) using a
$d_{x^{2}-y^{2}}$ gap,
$\Delta({\bf \hat k},T)=\Delta_{0}(T)\cos(2\phi)$ for a cylindrical Fermi
surface, the spectrum of
the phonon self energies can be written down in terms of complete
elliptical integrals $K$ and $E$ of the first and second kinds,
respectively. Taking screening into account and defining
$x=\omega/2\Delta_{0}$, we obtain for $T=0$,
\begin{equation}
\Sigma_{B_{1g}}^{\prime\prime \ sc.}=
\Sigma_{B_{1g}}^{\prime\prime}({\bf q}=0,\omega)=
{-4N_{F}g_{B_{1g}}^{2}\over{3\pi x}}
\cases{[ (2+x^{2})K(x)-2(1+x^{2})E(x)], & $x\le 1$
, \cr
x[ (1+2x^{2})K(1/x)-2(1+x^{2})E(1/x)],
& $x > 1$,}
\end{equation}
i.e., the $B_{1g}$ channel is not affected by Coulomb screening, while
\begin{equation}
\Sigma_{A_{1g}}^{sc.}=\Sigma_{A_{1g},A_{1g}}-
\Sigma_{A_{1g},1}^{2}/\Sigma_{1,1},
\end{equation}
with the spectral functions
\begin{eqnarray}
\Sigma_{A_{1g},A_{1g}}^{\prime\prime}({\bf q}=0,\omega)=
{-4N_{F}g_{A_{1g}}^{2}\over{15\pi x}}\ \ \ \ \ \ \ \ \ \ \ \ \ \ \ \ \ \ \ \ \
\ \ \ \ \ \ \ \ \ \ \ \ \ \ \ \ \ \ \ \ \ \ \ \ \ \ \ \ \ \ \ \ \ \ \
\nonumber \\
\times\cases{
[ (7-8x^{2}+16x^{4})K(x)-(7-12x^{2}+32x^{4})E(x)], & $x\le 1$, \cr
x^{4}[ (32-28/x^{2}+11/x^{4})K(1/x)-(32-12/x^{2}+7/x^{4})E(1/x)], & $x > 1$,}
\end{eqnarray}
\begin{equation}
\Sigma_{A_{1g},1}^{\prime\prime}({\bf q}=0,\omega)=
{-2\sqrt{2}N_{F}g_{A_{1g}}\over{3\pi x}} \cases{
[ (1+2x^{2})K(x)-(1+4x^{2})E(x)], & $x\le 1$, \cr
(1/x)[ (4-1/x^{2})K(1/x)-(4+1/x^{2})E(1/x)], & $x > 1$,}
\end{equation}
and
\begin{equation}
\Sigma_{1,1}^{\prime\prime}({\bf q}=0,\omega)=
{-2N_{F}\over{\pi x}}
\cases{[ K(x)-E(x)],& $x\le 1$ , \cr
x[ K(1/x)-E(1/x)], & $x > 1$.}
\end{equation}
The response functions for finite $T$ are obtained simply
by multiplying Eqs. (8) and (10-12) by the factor $\tanh(\omega/4T)$.
The partial screening of the $A_{1g}$ channel by long-range Coulomb
forces comes from the observation that the square of the energy gap enters
into the response function in Eq. (6). For the case of $d_{x^{2}-y^{2}}$
pairing symmetry,
the energy gap squared contains a term which transforms according to $A_{1g}$
symmetry which leads to a
mixing of the $L=0$ and $L=4$ $A_{1g}$ basis functions.
This corresponds to partial ``transverse screening'' of the $A_{1g}$
vertex\cite{us}.

The corresponding real parts were obtained via Kramers-Kronig analysis
and are plotted together with the imaginary parts in Figure 1 for the
$B_{1g}$ and screened $A_{1g}$ channels.
We see that the peak in the imaginary part of the self-energy
(which determines the linewidth of the phonon) lies at different frequencies
$\omega_{peak} \sim 2\Delta_{0}(T)$ and $1.2\Delta_{0}(T)$
for the $B_{1g}$ and $A_{1g}$ channels, respectively.
This is a consequence of the angular averaging which
couples the gap and $e-ph$ vertex, and leads to constructive
(destructive) interference
under averaging if the vertex and the gap have the same (different) symmetry.
Similar behavior for the electronic contribution to Raman scattering
led to the reasoning that the symmetry which shows the highest peak position
gives an unique indication of the predominant symmetry of the gap\cite{us}.
The symmetry dependence is also manifest in the low frequency behavior,
which can be written as
\begin{eqnarray}
\Sigma^{\prime\prime}_{B_{1g}}(\omega \rightarrow 0) =
3N_{F}g_{B_{1g}}^{2}x^{3}/4+O(x^{5}), \nonumber \\
\Sigma_{A_{1g}}^{\prime\prime}(\omega \rightarrow 0) =
N_{F}g_{A_{1g}}^{2}x+O(x^{3}),
\end{eqnarray}
i.e., the spectrum of the self energy rises slower in the
$B_{1g}$ channel than the $A_{1g}$ channel\cite{comp}.
The power laws are a signature of an energy gap which vanishes on lines on
the Fermi surface, but the channel dependence of the exponents are
unique to a $d_{x^{2}-y^{2}}$ pair state. These channel-dependent power-laws
have been observed in the electronic contribution to Raman scattering in
BSCCO, YBCO, and double and triple layer Thallium cuprates which constitute
strong evidence for a $d-$wave gap of this symmetry as
opposed to $d_{xy}, d_{xz}$ or  $d_{yz}$ symmetry, which also have
nodes on lines on the Fermi surface\cite{us}.

The real parts of the self energies (which determines the frequency
renormalization) show a change of sign near
the peak in the imaginary part.
While the $B_{1g}$ channel shows a mild frequency dependence away from the
peak maximum and then a rapid change of sign at the peak, the $A_{1g}$ channel
shows a smooth crossover from negative to positive values, with a change of
sign that occurs at a frequency which is slightly greater than the peak
maximum in the imaginary part.  Thus a phonon of $A_{1g}$ symmetry which lies
at energies below $2\Delta_{0}(T)$ can become hardened as opposed to softened.
We immediately can draw the conclusion that phonons
of the same frequency will show qualitatively
different behavior in different channels as a consequence of their symmetry.
Therefore, careful attention must be paid to symmetry before an analysis of
the gap can be made by locating the point where phonon softening or
hardening occurs.

\section{Temperature Dependence}
We now investigate the temperature dependence of the phonon lineshape.
The ${\bf q}=0$ spectral function, Eq. (6), vanishes at $T_{c}$
due to the lack of particle-hole continuum for pair
creation. This term thus always predicts phonon broadening compared to the
normal state below T$_{c}$ for a gap with nodes. However, the term
responsible for the normal metal self energy (due to, eg., finite
momentum transfer or anharmonic decay) will be affected by superconductivity
due to the reorganization of the density of states as the gap opens
up. In order to recover the normal metal lineshape at
$T_{c}$, one must use finite q (or impurity scattering \cite{me,zc}) to
generate
the additional term $\delta \Sigma$ which does not vanish at $T_{c}$.
We now generalize the result to finite $q$
for anisotropic gaps.
For finite $q$, the
spectrum of $\delta\Sigma$ at finite temperatures is given by
$\delta\Sigma^{\prime\prime}(q,T,\omega)=\Theta(v_{F}q-\omega)\langle
\mid g_{\bf \hat k}^{\gamma}\mid^{2}F({\bf \hat k},\omega)\rangle$ with
\begin{eqnarray}
F({\bf \hat k},\omega)={N_{F}\pi^{2}\over{2 v_{F}q}}
\int_{\mid\Delta({\bf \hat k})\mid}^{\infty}dE \Bigl\{[f(E)-f(E+\omega)]
{E(E+\omega)-\mid\Delta({\bf \hat k})\mid^{2}\over{\sqrt{E^{2}-\mid\Delta({\bf
\hat k})\mid^{2}}\sqrt{(E+\omega)^{2}-\mid\Delta({\bf \hat k})\mid^{2}}}}
\nonumber \\
+\Theta(E-\mid\Delta({\bf \hat k})\mid -\omega)[f(E-\omega)-f(E)]
{E(E-\omega)-\mid\Delta({\bf \hat k})\mid^{2}\over{\sqrt{E^{2}-\mid\Delta({\bf
\hat k})\mid^{2}}\sqrt{(E-\omega)^{2}-\mid\Delta({\bf \hat k})\mid^{2}}}}
\Bigr\},
\end{eqnarray}
where $f$ is a Fermi function. The Theta function $\Theta(v_{F}q-\omega)$
restricts the
frequency shift due to phase space consideration, reflecting that the
region of the particle-hole continuum vanishes for small wavenumbers as a
consequence of momentum conservation. Since $v_{F}q \ll \Delta$ in the
cuprate materials and also in A-15 materials, this term will only contribute
to the self energy for phonons of very small energy. However, it has been
shown for $s-$wave superconductors \cite{me} that the incorporation of impurity
scattering removes the phase-space restriction due to the lifting of
momentum conservation and $\delta\Sigma$ contributes for all frequencies.
While incorporating impurity scattering remains beyond
the scope of the present treatment, we remark that it is expected
that a similar
consideration for the case of $d-$wave superconductors would also lead
to the contribution of $\delta\Sigma$ for all frequencies.  This remains
to be explored\cite{tpd}.

In the limit of small frequencies $(\omega << T)$, we obtain the simple
result
\begin{equation}
\delta\Sigma^{\prime\prime}(q,\omega<<T)=\omega {2N_{F}\pi^{2}\over{v_{F} q}}
\langle
{\mid g_{\bf \hat k}\mid^{2}\over{e^{\mid\Delta({\bf \hat k})\mid/T}+1}}
\rangle \Theta(v_{F}q-\omega).
\end{equation}
Similarly, at $T_{c}$, Eq. (14) recovers
$\delta\Sigma^{\prime\prime}(q,\omega<<T_{c})=\Theta(v_{F}q-\omega)\omega
{N_{F}\pi^{2}\over{v_{F} q}}
\langle\mid g_{\bf \hat k}^{\gamma}\mid^{2}\rangle$ and
thus the ratio of the low frequency response in the superconducting state
to that of a normal metal at $T_{c}$ is given by
\begin{equation}
{\delta\Sigma^{\prime\prime}(q,\omega<<T)\over{\delta\Sigma^{\prime\prime}
(q,\omega<<T_{c})}}=2 {\langle \mid g_{\bf \hat k}^{\gamma}
\mid^{2} f(\mid\Delta({\bf \hat k})\mid)
\rangle\over{\langle \mid g_{\bf \hat k}^{\gamma}\mid^{2}\rangle}}.
\end{equation}
This shows how the redistribution of the density of states below $T_{c}$ to
higher energies as the gap opens up leads to a reduction of the
decay channels available to particle-hole creation and a net decrease
in the phonon linewidth.
In isotropic superconductors, the Fermi function can be pulled
out of the average and the resulting expression is independent of
phonon symmetry. However, once the gap is anisotropic, there exists coupling
between the vertex and the the gap which leads to a symmetry dependent
result.

Using a weak coupling expression for the temperature dependence of the
energy gap $(2\Delta_{0}/k_{B}T_{c}=5.1252)$,
we numerically evaluate Eq. (16) while taking screening into account.
The results are plotted in Fig. 2 as a function of $T/T_{c}$
for a $d_{x^{2}-y^{2}}$ energy gap compared
to a BCS isotropic gap.  The low temperature behavior is given by a
power-law in $T$ for all channels for the $d-$wave case while the
ubiquitous exponential dependence in $T$ is seen for all channels in the
$s-$wave case. The power-law behavior for the $d-$wave case is channel
dependent, with exponents identical to those of Eq. (13), in the
sense that $\omega$ can be replaced by $T$.
What is remarkable is that the fall off of the
Fermi function at low temperatures is quite slow in those channels which are
orthogonal to the symmetry of the gap, with the notable example of the
$A_{1g}$
channel, which shows a residual broadening at $T/T_{c}=0.3$ of roughly 20
percent of that of the normal state.
This was argued in the case of electronic Raman scattering to be
further evidence for an energy gap in the cuprate materials
which has predominantly $B_{1g}$ character, due to the observation that a
gap opens up quickly in the $B_{1g}$ channel compared to $A_{1g}$ and
others which have been probed via Raman scattering\cite{us}.

For higher frequencies $\omega>T$, we have evaluated Eq. 14 directly. The
results
are quite similar to those of Fig. 2 for all frequencies $\omega$ up
to roughly $4\Delta$, but then at higher frequencies
all channels eventually display a linear $T$
dependence (i.e., the behavior of the normal state)
for energy scales much greater than the gap energy.

\section{entire spectral function and role of satellites}

In this section we
consider the entire phonon spectral function Eq. (1), paying
particular attention to the role of satellites which arise due to
$e-ph$ coupling.  The role of
satellites have not been explored in anisotropic superconductors.
As is well known for
BCS superconductors, satellites appear in the phonon spectral function for
all frequencies of the optical phonon, but have the greatest residue for
phonons near twice the gap edge. In the BCS case, impurities wipe out the
satellite peak\cite{me},
explaining  why they have yet to be definitively observed
in conventional A-15 superconductors. In the absence of impurities
however vastly
different lineshapes can be obtained due to the interference
of the satellites.

Using a gap of $d_{x^{2}-y^{2}}$ symmetry and working specifically at
$q=0$, Eq. (6), we find that the satellites are
present in anisotropic superconductors as well due to the
fact that the real part of the denominator of Eq. (1) has two
zeroes for any value of $\omega_{0}$ - one at the renormalized phonon
frequency and the other at the satellite position. The satellite becomes
more pronounced the closer the optical phonon is to the peak position of the
self energy spectrum as in the BCS case and interferes with the
phonon. Therefore, for the case of
a phonon located below the spectral maximum, where the satellite peak
is observable only at T=0 for large
$e-ph$ coupling, as the gap decreases on
approaching T$_{c}$ the satellite will be made to pass through the phonon
position and will be subsequently distorted.
This is shown in Fig. 3 for a $B_{1g}$ phonon
($\omega_{0}/\Delta_{0}(T=0)=1.0, g_{B_{1g}}^{2}N_{F}/\Delta_{0}(T=0)=0.1$)
for the
temperatures indicated. The phonon lineshape is drastically affected by
the satellite which takes spectral weight away from the phonon when
the peak of the spectral function is close to the phonon position.
The phonon linewidth grows as the peak of the spectrum moves up in
energy with decreasing $T$ and is hardened. The linewidth and frequency
shift reaches a maximum when the peak and phonon position coincide and
then the linewidth decreases and the phonon softens as $T \rightarrow
T_{c}$.

\section{Conclusions and Comparison with Experiment}

In this Section we combine the previous results and
examine the phonon linewidth as a function of
temperature for the case of two phonons which lie at
approximately $285, 340$ cm$^{-1}$ for the $B_{1g}$ channel
and $464, 500$ cm$^{-1}$ for the $A_{1g}$ channel in BSCCO and YBCO,
respectively.
Using our previous
fits to the electronic Raman scattering in BSCCO, we obtained a value of the
energy gap at $T=0$ to be $\Delta_{0}(T=0)=287$ cm$^{-1}$\cite{us}.
Therefore the normalized optical phonon frequency is given by
$\omega_{0}/\Delta_{0}(T=0) \sim 0.99, 1.62$ for the $B_{1g}, A_{1g}$
phonon, respectively in BSCCO, while for YBCO the ratio is expected to be
slightly higher. We can immediately make the following statement.
Since the interference effects of the phonon with the satellite peak
can only occur for a phonon which is located at $T=0$ {\it below}
the peak in the imaginary part of the self energy, there should be no
interference effects on the $A_{1g}$ phonon since it lies {\it above}
the peak in the spectrum at $T=0$. Thus its renormalization should be a
monotonic function of temperature. However that is not the case for the
$B_{1g}$ phonon. Anomalous behavior of the $B_{1g}$
renormalization as seen in Fig. 3 arises
due to the interference between the phonon and the rapid
rise of the self energy near $2\Delta_{0}(T)$, which passes through the
phonon frequency at T/T$_{c} \sim 0.9$.
Another remark is in order. In order to make an accurate fit to the data,
the magnitude of the coupling constant needs to be addressed. As we have
seen in Section IV, it controls the strength of the satellite and
its subsequent effect on the phonon lineshape. Little is known about
the coupling constant \cite{hung} and thus we can only make general
statements on the behavior of the phonons. The magnitude of the effect
cannot be predicted.

Inspecting Fig. 1, the ${\bf q}=0$ part of the
self energy, at each phonon frequency,
we see that the $B_{1g}$ phonon is broadened and softened
at $T=0$ compared to $T_{c}$ while the $A_{1g}$ phonon is broadened but
lies right at the point where the real part is changing sign.
This term most accurately describes what is seen in the phonons in YBCO.
A rapid rise of the $B_{1g}$
phonon linewidth below $T_{c}$ has been seen\cite{irwin,klein}, reflecting the
interference of the peak in the self energy and the phonon,
(see Eq. (6) and Fig. 3). The $B_{1g}$ phonon additionally is mostly
softened to lower frequencies. For the $A_{1g}$ phonon, the frequency
shift is seen only to slightly higher frequencies. This is due to
the real part of the self energy crossing zero at the phonon position.
However, the $A_{1g}$ phonon narrows just below $T_{c}$, reaches a
minimum width and then broadens at lower temperatures. This could be a
result of a competition between the effects of $\Sigma$ and $\delta\Sigma$,
but without information concerning the magnitude of the coupling
constant, finite momentum transfer or impurity scattering
this remains an open question.

In addition, we have the contribution arising from non-zero $q$, Eq. (14),
which indicates that this contribution to the self energy is reduced
compared to its value at $T_{c}$ and decreases
as $T^{3}$ and $T$ for low temperatures in the $B_{1g}$ and
$A_{1g}$ channels, respectively (see Fig. 2). This term most accurately
describes the experiments in BSCCO. The linewidth of both the $B_{1g}$
and $A_{1g}$ phonons decrease monotonically with temperature, which
points to the lack of contribution coming from satellite effects. The
linewidth decrease in the $B_{1g}$ channel can be fit with a $T^{3}$
dependence while a term linear in $T$ can be fit to the $A_{1g}$
phonon which is the same dependence as in the normal state. Both behaviors
are consistent with Eq. (16). This is to be compared with
the predictions of an isotropic $s-$wave theory
lines, which are identical for each channel ($\sim e^{-\Delta/T}$).
The theory for a gap of $B_{1g}$ character shows a marked symmetry
dependence resulting from the interplay of gap and vertex symmetry. While it
is arguable whether an exponential temperature dependence can also fit
the $B_{1g}$ data \cite{leach}, the lack of change of the exponent of
the $A_{1g}$ phonon's linewidth is a direct consequence of the energy gap
anisotropy. Therefore it will appear that the $A_{1g}$ phonon will be
unaffected by superconductivity. Since the $B_{1g}$ phonon has the same
symmetry of a $d_{x^{2}-y^{2}}$ gap, it's linewidth will show the greatest
change due to the onset of superconductivity.

In conclusion,
within the accuracy of the experiments, we have seen that the changes in the
phonon lineshape as a function of symmetry can be explained with a choice
of the gap which has (at least predominantly) $B_{1g}$ character,
supporting recent comparisons made on electronic Raman scattering on
BSCCO. However, without knowledge of the magnitude of the coupling
constant, the importance of finite $q$ and satellite effects remains
an open question. Of course
other choices of gaps which have a small but finite minimum value, eg.
anisotropic $s$-wave or $s+id$ would give similar results to the
$d_{x^{2}-y^{2}}$ choice for the gap. Both the electronic and phonon
contributions to Raman scattering below $T_{c}$ can be explained by simply
invoking the symmetry of the vertex which couples to the symmetry of the
gap.  More detailed experiments would be extremely useful to pin down the
magnitude of the $e-ph$ vertex and subsequently the role of the satellites,
and the role of impurities and the mechanism and magnitude of the
coupling remains to be explored\cite{hung,tpd}.

\acknowledgements
I am indebted to Drs. D. Einzel, A. Virosztek and A. Zawadowski
for many valuable
discussions and a critical reading of
the manuscript. I would also like to thank D. Leach for discussions and for
providing me with the data discussed in the text prior to publication.
Similarly, I would also like to acknowledge helpful
suggestions and discussions from Drs. R. Hackl, E. Nicol, B. Stadlober,
and G. Zimanyi. I would like to acknowledge the hospitality of A.
Virosztek and A. Zawadowski and their colleagues at the
Research Institute for Solid State Physics
of the Hungarian Academy of Science and the Institute of Physics of the
Technical University of Budapest, where part of this work was completed.
This work was supported by N.S.F. grant 92-06023 and the American Hungarian
Joint Grant number NSF 265/92b.

\figure{Real and imaginary parts of the phonon self energy for the $B_{1g}$ and
$A_{1g}$ channels for a cylindrical Fermi surface. Magnitude of the vertices
are set equal to one.}
\figure{Temperature dependence of the $\omega \rightarrow 0$ imaginary part
of the self energy in a $d_{x^{2}-y^{2}}$ and isotropic BCS superconductor
compared to the normal state.}
\figure{Phonon spectral function for a $B_{1g}, (\omega_{0}=\Delta_{0}(T=0),
g_{B_{1g}}^{2}N_{F}/\Delta_{0}(T=0)=0.1)$,
phonon in a superconductor with $d_{x^{2}-y^{2}}$ pairing symmetry for
various values of $T/T_{c}$ as indicated in upper part of Figure.}
\end{document}